\documentstyle[epsfig]{new}

\begin{document}

\title{Model Independent Analysis of the Solar Neutrino Data
\footnote{Talk presented by H. Nunokawa at the symposium, 
"New Era in Neutrino Physics", Tokyo Metropolitan University, 
Japan, 11-12 June 1998, to be published in the proceedings.}}

\author{Hisakazu MINAKATA \\
{\it Department of Physics, Tokyo Metropolitan University,\\
1-1 Minami-Osawa Hachioji-shi, Tokyo 192-0397, Japan, \\
minakata@phys.metro-u.ac.jp}\\
Hiroshi NUNOKAWA\\
{\it Instituto de F\'{\i}sica Gleb Wataghin\\
Universidade Estadual de Campinas - UNICAMP\\
P.O. Box 6165, 13083-970 Campinas SP Brazil,\\
nunokawa@ifi.unicamp.br}}

\maketitle

\section*{Abstract}
We perform an updated model-independent analysis using all 
the latest solar neutrino data, including the one coming 
from remarkably high statistics SuperKamiokande experiment.  
We confirm that the astrophysical solutions to the solar neutrino 
problem are extremely disfavored.  
We also present a new way of illuminating the suppression pattern 
of various solar neutrino flux, which indicates that the strong 
suppression of $^{7}$Be neutrinos is no more true once the neutrino 
flavor conversion is taken into account. 

\section{Introduction}


The solar neutrino problem [2] is now established 
essentially independent of any details of the standard solar 
models (SSM). 
The so called model-independent analysis was performed by several 
authors [{3},{10},{4},{19},{16},{24},{5},{17},{13}] 
which revealed that the $^{7}$Be neutrinos must be 
strongly depleted compared to the SSM prediction.
{}From these analyses one can conclude that the solar neutrino 
problem cannot be explained by astrophysical mechanisms unless 
some assumptions in the standard electroweak theory and/or solar 
neutrino experiments are grossly incorrect.

In this work, we repeat the model-independent analysis using the 
latest solar neutrino data as well as the new updated theoretical 
prediction by Bahcall and Pinsonneault (BP98) [{7}]. 
Our analysis will indicate that astrophysical solution such 
as the low-$T$ model [{2}] is convincingly excluded by the 
present data. 
We also present a new way of illuminating the suppression 
pattern of various solar neutrino flux originated from 
different fusion reactions in a less model-dependent fashion.  
We will observe that the statement of the missing $^{7}$Be neutrinos
is no more true in the presense of neutrino flavor conversion.  

\begin{table}[h]
\label{tab:data}
\caption{Observed solar neutrino event rates used in this analysis and 
corresponding theoretical predictions [{7}]. 
The quoted errors are at $1\sigma$.}
\vspace{.5pc}
\begin{tabular}{ccccc} \hline
Experiment 	& Data~$\pm$(stat.)~$\pm$(syst.)&Ref.
& Theory \protect [{7}] & Units \\ \hline
Homestake	& $ 2.56 \pm 0.16 \pm 0.15$        & \protect [{11}] &
	$7.7^{+1.2}_{-1.0}$    & SNU					\\
SAGE		&$69.9^{+8.0}_{-7.7}{}^{+3.9}_{-4.1}$& \protect [{14}]&
	$129^{+8}_{-6}$        & SNU					\\
GALLEX		& $76.4 \pm  6.3^{+4.5}_{-4.9}$     & \protect [{18}]&
	 $129^{+8}_{-6}$       & SNU					\\
SuperKam	&$2.44 \pm0.05{}^{+0.09}_{-0.07}$& \protect [{23}]&
 	$5.15^{+0.98}_{-0.72}$  & \hglue -0.2cm $10^6$ cm$^{-2}$s$^{-1}$ \\ \hline
\end{tabular}
\end{table}

\section{The Data}

The latest solar neutrino data, including the $^{8}$B neutrino flux 
measured during 504 days in the SuperKamiokande experiment [{23}],  
which will be used in our analysis are tabulated in Table 1. 
{}From these data,  by adding the statistical 
and the systematic errors quadratically, we obtain,
\begin{eqnarray}
S^{obs}_{Cl}&=&2.56  \pm 0.23 \ \ \ \mbox{SNU},\\
S^{obs}_{Ga} & = & 72.4 \pm 6.6 \ \ \ \mbox{SNU},\\
S^{obs}_{SK}  & = & (2.44 \pm 0.10) \ \times \ 10^6 \ \mbox{cm}^{-2}\mbox{s},
\label{obscombined}
\end{eqnarray}
where, to be conservative, we always take the larger values of 
statistical and systematic errors, whenever errors are asymmetric, 
in each experiment before we combine. 

Our analysis in the present work is based only upon the total rate 
of each experiment, and the information of the energy spectrum of 
$^8$B neutrinos obtained by SuperKamiokande experiment, is not 
taken into account. 
Thus, it is to illuminate the global features of the suppression 
of the solar neutrino spectrum. 
%


\section{Model-Independent Analysis}

The fundamental assumptions in our analyses are as follows:
(i) The sun shines due to the nuclear fusion reactions from which  
and only from which the solar neutrinos come, 
(ii) the relevant reactions responsible for creating 
neutrinos in the sun are assumed to be those postulated 
in the SSM, 
(iii) the sun is quasi-stable during the time scale of 0.1-1 million years.

These assumptions (i) to (iii) imply that the solar neutrino flux 
generated by various nuclear fusion reactions must obey the luminosity 
constraint [{21},{22},{15},{24},{5}], 
\begin{equation}
\label{solarlumi}
\frac{L_\odot}{4\pi R^2} 
= \sum_\alpha \left(\frac{Q}{2}-\langle {E} \rangle_\alpha \right) 
\Phi(\alpha)
\end{equation}
where 
$L_\odot = 3.844 \times 10^{33}$ (erg/s),  
$R$ = 1 A.U. ($1.469 \times 10^{13}$ cm), 
$Q$ = 26.73 MeV is the energy release when one $^4$He is created, 
$\langle {E} \rangle_\alpha$ and 
$\Phi(\alpha)$, ($\alpha = pp$, $^7$Be, $^8$B,...),
are the average energy loss by neutrinos and the neutrino flux, 
respectively. 

By normalizing the neutrino flux to those of the SSM of BP98 [{7}] as,
\begin{equation}
\label{norm}
\phi^\alpha = \frac{\Phi(\alpha)}{\Phi(\alpha)_{SSM}},   \ \ 
(\alpha = pp, ^7\mbox{Be}, ^8\mbox{B},...), 
\end{equation}
the luminosity constraint is simply given by,
\begin{equation}
\label{lumi}
1 = 0.907 \phi^{pp} + 0.0755  \phi^{^7 Be} + 4.97 \times 10^{-5}\phi^{^8 B},
\end{equation}
where we have neglected the contribution from  CNO and $pep$ neutrinos.

In this section we make the following two more specific assumptions 
in addition to the fundamental assumptions (i) - (iii): 
(1) The energy spectra of the solar neutrinos are not modulated, 
(2) neutrino flavor transformation does not occur 
after neutrinos are created in the sun until they reach the detectors. 
Then the luminosity constraint is effective, and the flux $\Phi$ is 
the actual flux to be detected by the terrestrial detectors. 

The expected solar neutrino signal to the $^{37}$Cl, $^{71}$Ga 
and SuperKamiokande solar neutrino experiments are given in terms 
of neutrino flux by,
\begin{eqnarray}
\label{theorycl}
S^{th}_{Cl}&=&5.9 \phi^{^8 B} + 1.15 \phi^{^7 Be} 
 \ \ \ \mbox{SNU},\\
\label{theoryga}
S^{th}_{Ga} & = & 12.4 \phi^{^8 B} + 34.4 \phi^{^7 Be} 
+ 69.6 \phi^{pp} \ \  \ \ \mbox{SNU},\\
\label{theorysk}
R^{th}_{SK}&=& \phi^{^8 B},
\end{eqnarray}
where we have neglected the contribution from the $pep$ and the CNO 
neutrinos because the inclusion of them does not affect our 
conclusion in this section.

Using Eqs. (\ref{theorycl}-\ref{theoryga}) as well as the 
observed solar neutrino data summarized in Table I we perform a 
simple $\chi^2$ analysis. 
After eliminating $\phi^{pp}$ from (8) by 
using the luminosity constraint we freely vary the two 
flux, $\phi^{^8 B}$ and $\phi^{^7 Be}$ and compute the $\chi^2$. 
The minimum $\chi^2$ is reached when $\phi^{^7 Be}$ takes a 
negative value but we impose the condition 
$\phi^{^7 Be}\geq 0$ to be physically meaningful. 

\begin{figure}[ht]
\hglue 0.8cm
\epsfig{file=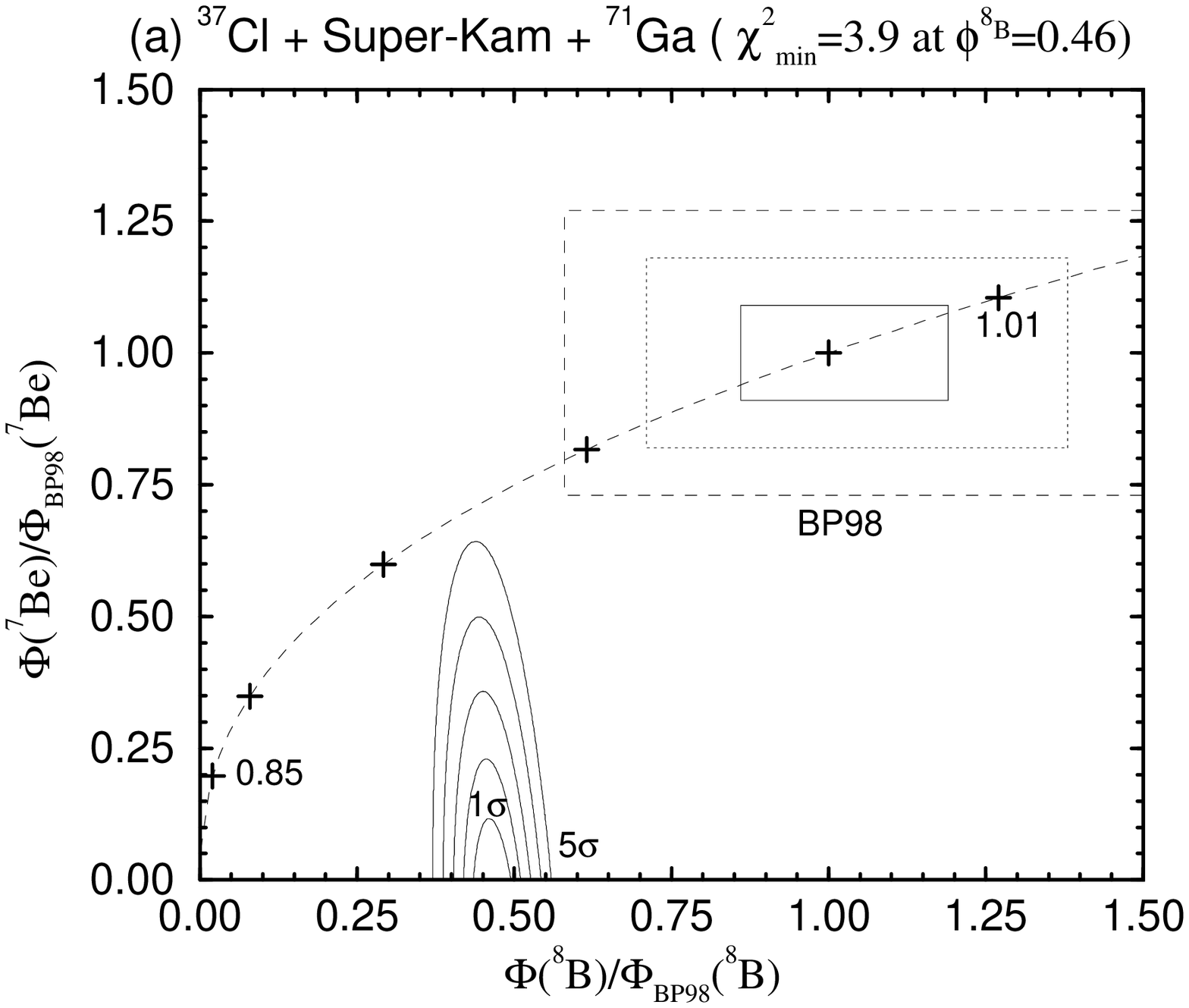,width=6.5cm}
\vglue -5.46cm \hglue 7.7cm \epsfig{file=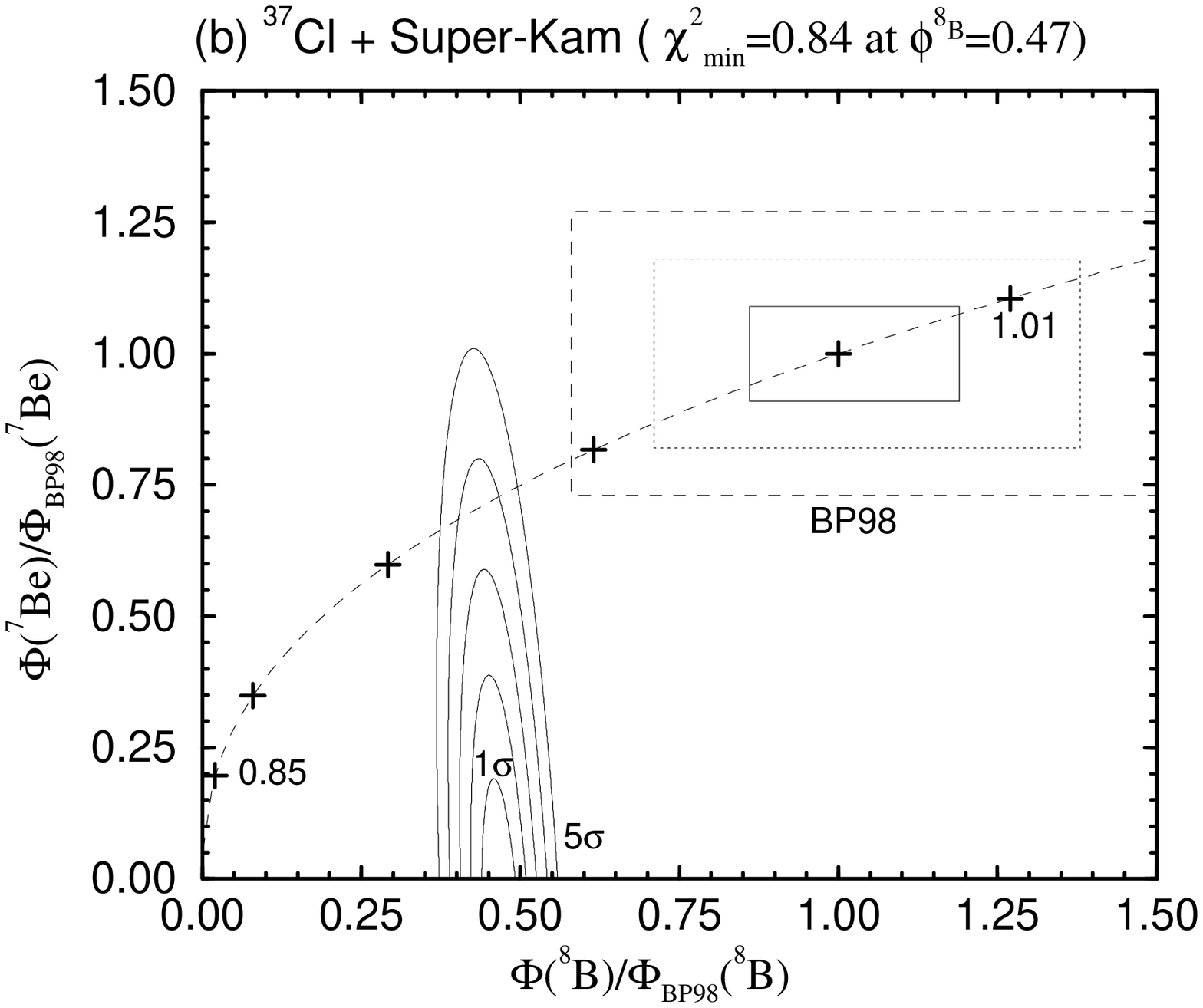,width=6.5cm}

\hglue 0.8cm
\epsfig{file=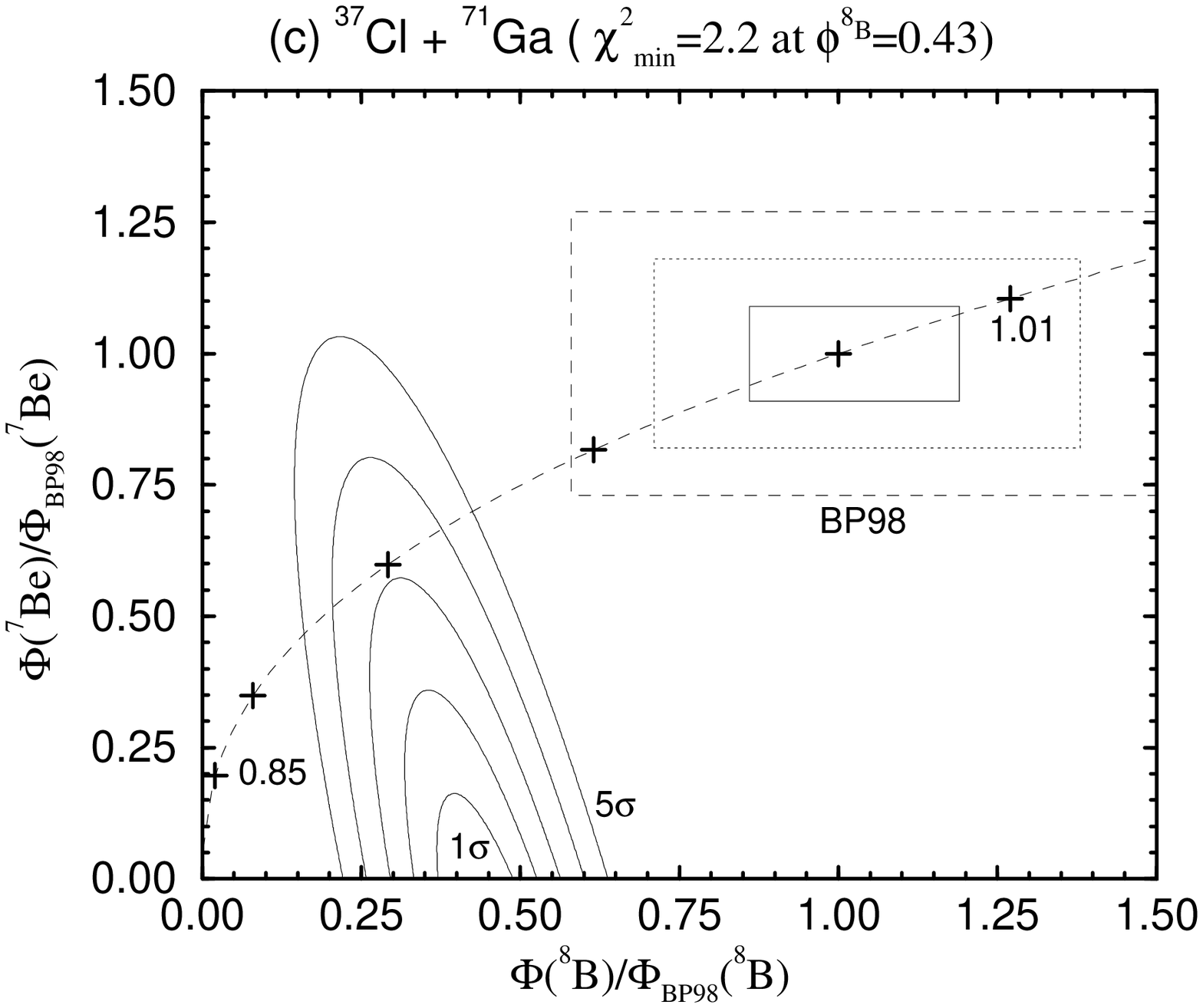,width=6.5cm}
\vglue -5.36cm \hglue 7.7cm \epsfig{file=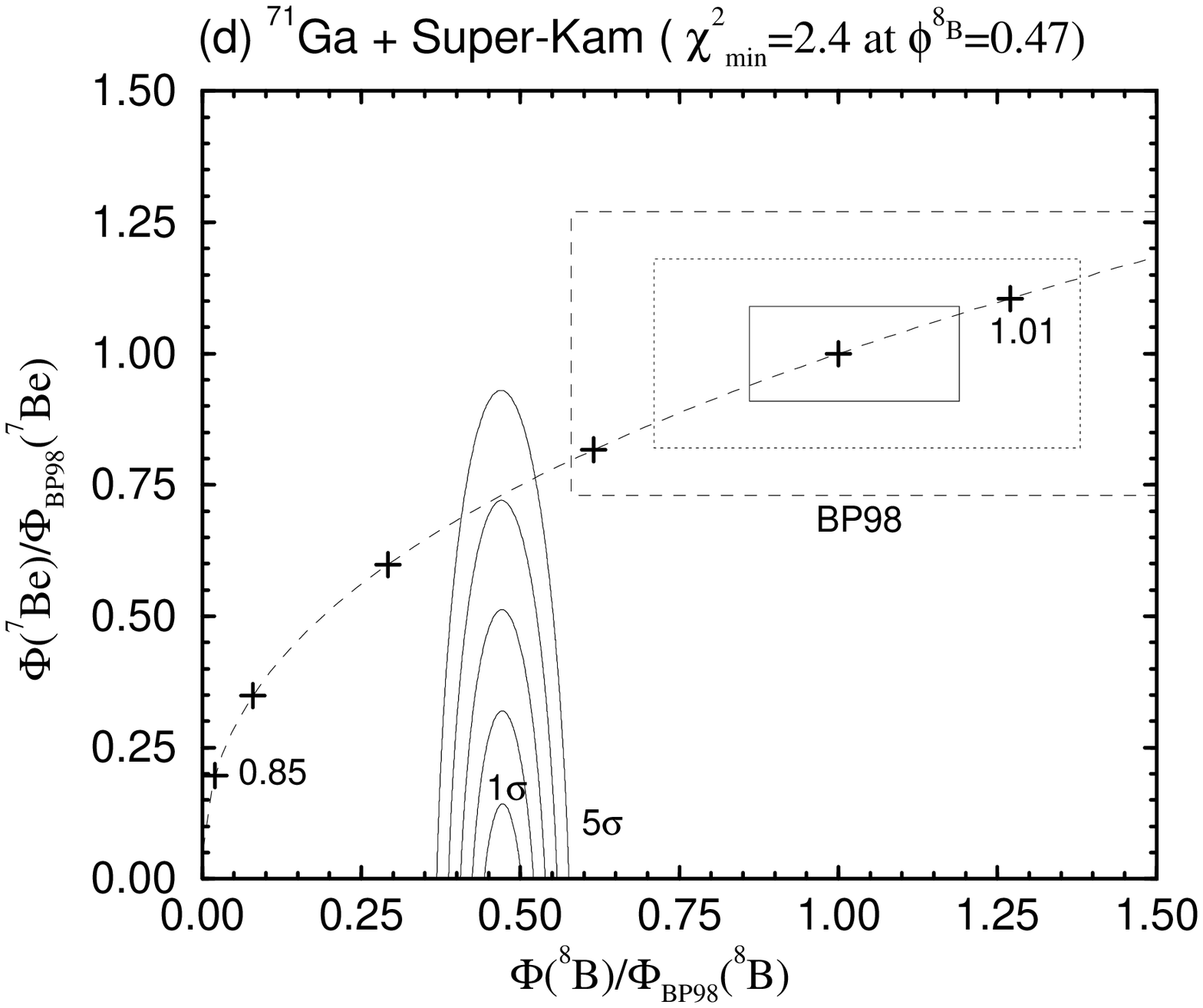,width=6.5cm}
\vglue -0.2cm
\caption{
Contour plot of the $\chi^2$ values in the 
$\Phi_{^8B}-\Phi_{^7Be}$ plane for different combinations
of the solar neutrino experiments. The solid curves 
correspond to 1 $\sigma$ to 5 $\sigma$, with step size 1,  
from inside to outside. We also indicate the 1,2 and 3 $\sigma$ 
theoretical range predicted by BP98, by the solid, dotted and dashed 
lines, respectively. 
Along the dashed curve, $\phi^{^7Be} = (\phi^{^8B})^{10/24}$,  
the crosses indicate, from left to right,  
the point where the central temperatures are 0.85, 
0.9, 0.95, 0.98, 1 and 1.01 with respect to the 
prediction by the SSM. 
}
\label{fig:parke1}
\end{figure}

We plot in Fig. 1 
the contours of $\chi^2$ corresponding to 1$\sigma$, 2$\sigma$, ... 5 
$\sigma$, for two free parameters. 
We also plot in Fig. 1 the curve, $\phi^{^7Be} = (\phi^{^8B})^{10/24}$ 
which corresponds to the case where the central temperature of 
the sun $T_c$ is varied freely, but keeping the relationship between 
the two fluxes as [{6}], $\phi^{^7Be} \propto T_c^{10}$ and 
$\phi^{^8B} \propto T_c^{24}$. 

As in the previous analyses [{15},{10},{4},{19},{16},{24},{17}] 
$\chi^2_{min}$ is achieved at vanishing $^7$Be flux 
not only in Fig. 1 (a) where all the experiments are taken into 
account but also in Fig. 1 (b-d) where only two experiments 
are included. 
By comparing Fig. 1, for example, with those of [{15}]
and by [{24}], we can clearly see that width of the contours 
has been greatly shrunk along the $\phi^{^8B}$ axis, 
indicating  how significantly the result is affected by 
the high statistics data from SuperKamiokande.

{}From Fig. 1 we see that solar neutrino data are in strong 
disagreement with the SSM prediction as also concluded in ref. [{8}]. 
We conclude from Fig. 1 that the standard solar model BP98 is ruled 
out by the current solar neutrino data at the significance level 
much higher than 5$\sigma$ under our fundamental assumptions (i)-(iii) 
and the additional ones (1)-(2). 
We have also confirmed that the astrophysical solution of the solar 
neutrino problem, such as low-$T$ model
is strongly disfavored by the data. 
We stress that our basic conclusions do not change even if 
we neglect one of the three types of experiments. 


\section{Suppression Pattern of Neutrino Flux implied by the 
current Solar Neutrino Data}

In this section we describe a new way of illustrating the suppression 
pattern of most relevant neutrino flux, required to explain the 
current solar neutrino experiments. 
We do this by taking into account the possibility of occurrence of 
either the active conversion, $\nu_e\to\nu_{\mu,\tau}$ or 
the sterile one $\nu_e\to\nu_s$ in between the solar core 
and the terrestrial detectors. 

To obtain global understanding of the suppression pattern
we propose to combine the $pep$ and CNO neutrinos into 
the $^{7}$Be neutrinos and denote them as the intermediate 
energy neutrinos. 
While it is more reasonable, in the context of model-independent analysis, 
to combine the $pep$ neutrinos with the ${pp}$'s because they are competing 
partners in the ${pp}$ I chain, here,  we combine the flux when their energy 
regions overlap.

We assume, in this section (except in Subsec. 4.3), that neutrino 
production rates from each source are the same as the ones predicted 
by the BP98 SSM. Then the expected signal in each experiment in the 
presence of neutrino conversion, $\nu_e\to\nu_{\mu,\tau}$, 
is given by, 
\begin{eqnarray}
\label{theorycl2}
S^{th}_{Cl}&=&5.9 \langle {P_B} \rangle + 1.83 \langle{P_I}\rangle
 \ \ \ \mbox{SNU},\\
\label{theoryga2}
S^{th}_{Ga} & = & 12.4 \langle{P_B}\rangle+ 46.9 \langle{P_I}\rangle
+ 69.6 \langle {P_{pp}} \rangle  \ \  \ \ \mbox{SNU},\\
R^{th}_{SK}&=& \langle {P_B} \rangle + r(1-\langle {P_B} \rangle), 
\label{theorysk2}
\end{eqnarray}
where $\langle {P_B} \rangle$, $\langle {P_I} \rangle$ and 
$\langle {P_{pp}} \rangle$ are the average survival probabilities for $^{8}$B, 
intermediate energy and $pp$ neutrinos, respectively. 
They are regarded as the average over the neutrino flux times the 
cross section, and as well as the detection efficiency in the case of 
the SuperKamiokande experiment. 
In eq. (\ref{theorysk2}) $r$ is essentially given by the ratio 
of the scattering cross section of $\nu_{\mu(\tau)}$ to that of 
$\nu_e$ off electron, i.e., 
$r \equiv {\langle {\sigma_{\nu_\mu e}} \rangle}/
{\langle {\sigma_{\nu_e e}} \rangle} \simeq 0.16$. 
When we consider the case (in Subsec. 4.2) 
where the $^8$B $\nu_e$'s are converted 
into some sterile states, 
we set $r=0$ in eq. (\ref{theorysk2}). 

In eqs. (12-14) we simply assume that the average survival probability 
for all the intermediate energy $pep$, CNO and $^7$Be neutrinos are 
the same and denote it as $\langle{P_I}\rangle$ so that the coefficient 
of $\langle{P_I}\rangle$ in eqs. (12) and (13) now includes the 
contribution not only from $^7$Be but also from $pep$ and CNO neutrinos 
(cf. eqs. (7) and (8)). 
Furthermore, we take, as an approximation,  
$\langle{P_i}\rangle$ ($i=$ $pp$, $I$, $^8$B) to be 
equal for all the experiments 
because the energy dependences of the flux times cross 
section (times the detection efficiency for the 
SuperKamiokande) are rather similar among 
different experiments, as first noticed
by Kwong and Rosen [{19}].

Other than these assumptions, we do not consider any specific mechanism 
of neutrino flavor transformation in this analysis but aim at illuminating 
global features of the modification of the solar neutrino spectrum. 
We try to determine the reduction rates of the flux of low, 
intermediate, and high energy neutrinos 
at the earth in such a way that the experimental data can be fitted. 
%

\begin{figure}[ht]
\vglue -3.8cm
\hglue -3.0cm
\epsfig{file=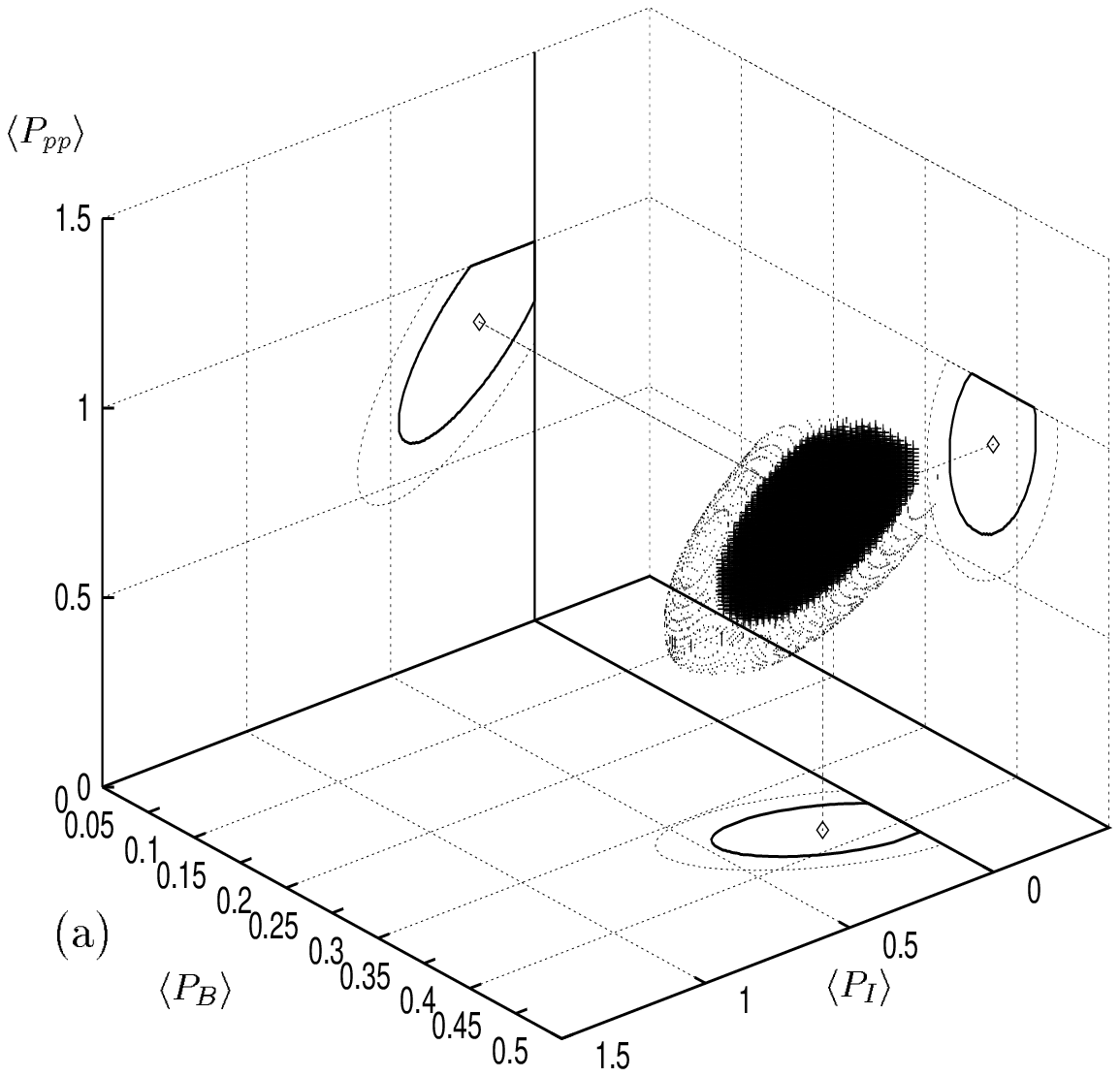,width=12.0cm}
\vglue -17.0cm \hglue 5.0cm \epsfig{file=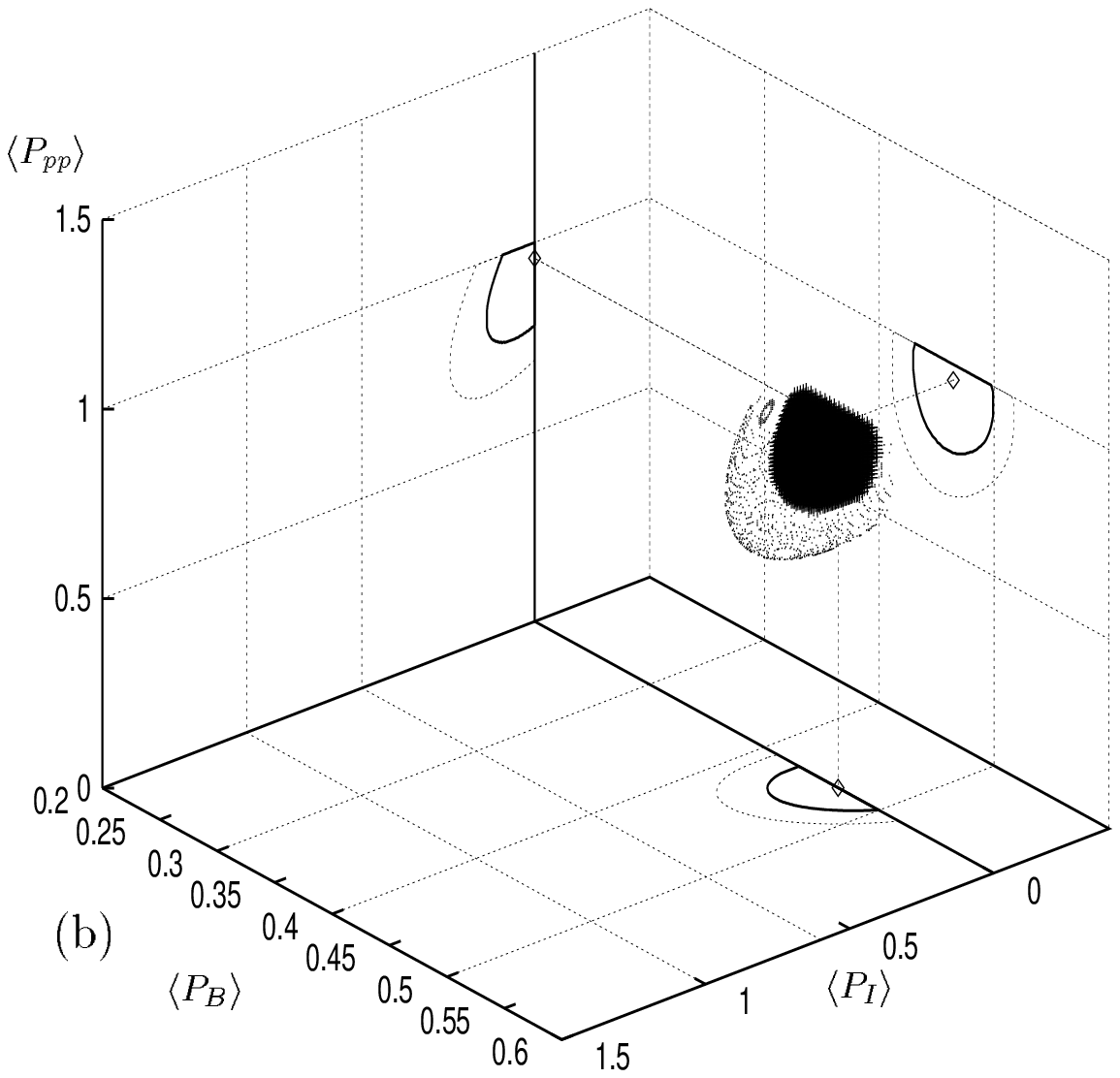,width=12.0cm}
\vglue -6.0cm
\caption{
Allowed range of neutrino flux determined by all the solar 
neutrino experiments with the condition $\chi^2=\chi^2_{min}+3.5$ 
(1$\sigma$) and 8.0 (2$\sigma$) 
(for three free parameters) assuming the neutrino conversion 
$\nu_e \to \nu_{\mu,\tau}$ or $\nu_s$. 
We show in (a) the active and in (b) the sterile conversions cases. 
The allowed ranges are projected into each plane, indicated by 
the solid curves (1$\sigma$) and the dotted curves (2$\sigma$). 
The best fitted reduction rates of the neutrino fluxes are, 
($\langle {P_B}\rangle$, $\langle {P_I} \rangle$, $\langle { P_{pp} } \rangle$) 
= (0.37,  0.19, 0.84) for active conversion with $\chi^2 \approx 0$, 
and =(0.47, 0, 0.96) for sterile conversion
with $\chi^2 \approx 0.8$. 
\label{fig:3dactive1} 
}
\vglue -0.5cm
\end{figure}
\subsection{The case of active neutrinos}

We present our results in Fig. 2 (a) for the case of active conversion.  
We note that $\langle {P_B} \rangle$ is determined most accurately, 
as expected from the large statistics of the SuperKamiokande  
experiment. On the other hands, the other two, $\langle {P_I} \rangle$ and 
$\langle {P_{pp}} \rangle$, have larger uncertainties at the present stage 
of the solar neutrino data. 
We stress that the proposed experiments such as Borexino 
[{1}], Hellaz [{20}] and Heron [{9}] are needed in order 
to determine the $^7$Be and $pp$ neutrino flux 
more accurately, especially if the conversion 
mechanism is unknown. 
We also tabulate the range of allowed values of the survival 
probabilities with their 1 $\sigma$ uncertainties in Table 2.

{}From Fig. 2 and Table 2 we can see that strong suppression of 
intermediate energy neutrinos, the one best fit by negative flux, 
is no more true when the neutrino 
flavor conversion is taken into account. 
This feature is in sharp contrast with the results of the model-independent 
analysis in Sec. 3 and of the flavor conversion into sterile neutrinos 
to be discussed below. 

We also note that the suppression rate of the intermediate-energy 
neutrinos depends rather sensitively on the presense or absence 
of the $pep$ and CNO neutrinos. If we ignore their contribution 
the best fit value of $\langle {P_I} \rangle$ 
(=$\langle {P_{Be}} \rangle$) becomes 
larger by a factor of 2 (see Table 2).

\begin{table}[h]
\caption[Tab]{The range of reduction rates of each neutrino flux 
with respect to the prediction by BP98 SSM 
implied by the solar neutrino data. 
We present the cases with (a) and without (b) 
pep and CNO contribution. 
}  
\vspace{.5pc}
\begin{center}
\begin{tabular}{cccc}\hline
Case & $\langle {P_B} \rangle$  & $\langle {P_I} \rangle$ 
& $\langle {P_{pp}} \rangle$   \\ \hline
Active (a)  &  $0.33-0.42$   & $0-0.46$  & $0.6-1$    \\ 
Active (b)  &  $0.33-0.42$   & $0-0.74$  & $0.55-1$ \\ \hline
Sterile (a)  &  $0.43-0.50$   & $0-0.16$  & $0.77-1$  \\ 
Sterile (b) &  $0.43-0.50$   & $0-0.26$  & $0.76-1$ \\ \hline
\end{tabular}
\end{center}
\label{tab:chi2}
\end{table}

\begin{figure}[ht]
\vglue -5cm
\centerline{
\epsfig{file=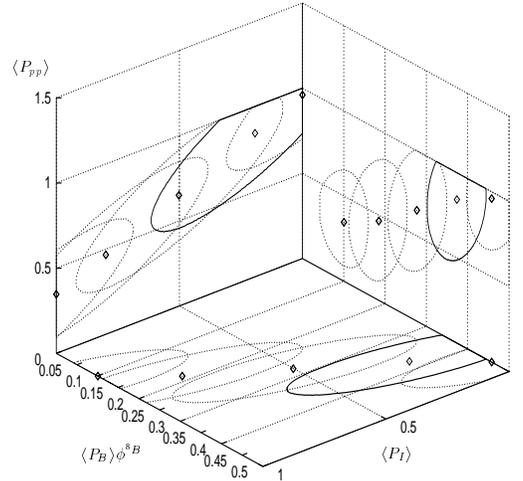,width=12.0cm}}
\vglue -6cm
\caption{
Two sigma allowed range of neutrino flux, projected into each plane, 
assuming the neutrino conversion $\nu_e \to \nu_\mu$ or $\nu_\tau$, 
for different values of $\phi^{^8B}$ are plotted by the dotted 
curves (except for the $\phi^{^8B}$ = 1 case).
The five curves in the each plane correspond, from left to right, 
to the case where $\phi^{^8B}$ = 2.5, 2.0, 1.5, 1.0 and  0.5. 
The corresponding best fitted reduction rates, indicated 
by open diamonds are, 
($\langle {P_B} \rangle \phi^{^8B}$, 
$\langle {P_I} \rangle$,  $\langle {P_{pp}} \rangle$) 
= (0.1, 1.0, 0.35), (0.18, 0.81, 0.46), (0.28, 0.50, 0.65), 
(0.37, 0.19, 0.84) and (0.46, 0.0, 0.96). 
\label{fig:3active2}}
\end{figure}


\subsection{The case of sterile neutrinos}

We next consider the case where the neutrinos are converted 
into sterile species. Since only the water-Cherenkov experiment 
can be sensitive to the difference between conversions into 
active and sterile neutrinos any change in our result from the 
active case solely comes from $^{8}$B neutrinos.  
The results for the sterile neutrino conversion is presented 
in Fig. 2 (b) and in Table 2. 
By comparing Fig. 2(a) and (b), 
we can clearly see that the stronger 
suppression of $^{7}$Be neutrinos is required than the case of 
active conversion in the case with $\phi^{^8B}=1$. 
We note that the best fit is obtained when the flux of 
intermediate energy neutrino is negative. 

\subsection{ Varying $^{8}$B flux}

Finally, we discuss the sensitivity of our results against 
the change in the neutrino flux from those of the SSM. 
Since the $pp$ neutrino flux is essentially fixed by the 
solar luminosity and also the $^{7}$Be neutrino flux are 
better determined compared to the $^{8}$B flux which is 
subject to the uncertainty of the nuclear 
cross section $S_{17}$, we only vary the $^{8}$B flux 
and examine the sensitivity of the required reduction rates 
against its change. 

We will perform this exercise only for the active neutrino conversion case
since for the sterile case the result presented in Fig. 2 (b) 
still holds if $\langle {P_B} \rangle$ is regarded as 
$\langle {P_B} \rangle\phi^{^8B}$ if we vary $\phi^{^8B}$, 
whereas for the active case, this is not true, because 
in the water-Cherenkov experiment, 
the event rate depends not only on $\phi^{^8B}\langle {P_{B}} \rangle$ 
but also on $\phi^{^8B}$ itself. 

In Fig. 3 we plot the allowed range of the reduction rates of 
the neutrino flux by (artificially) varying the $^{8}$B 
neutrino flux prediction of the SSM. 
The result of the exercise indicates that as the $\phi^{^8B}$ 
gets larger, preferred value of $\langle {P_I} \rangle$ becomes larger, while 
$\langle {P_{pp}} \rangle$ gets smaller as seen in the plot. 
This feature is consistent with one obtained by Smirnov in 
Table I of [{25}]. 

Let us note that the arbitrariness of the interpretation of 
which $\phi^{^8B}$ or $\langle {P_{B}} \rangle$ are changed from the standard 
theory can be removed if we combine the results of the 
SuperKamiokande and either one of charged or (preferable) neutral 
current data from the SNO experiment [{12}].
One can separately estimate the flux of $^{8}$B neutrinos and
the survival probability by combining these two experiments.


\section{Conclusions}


We have performed the updated model-independent analysis of the 
current solar neutrino data.
We confirmed, with current data of any two sets out of the three, 
the $^{37}$Cl, the $^{71}$Ga and the SuperKamiokande 
experiments that: 
(1) the SSM prediction can be convincingly rejected, and 
(2) the $^{7}$Be neutrinos is strongly suppressed unless
$^{8}$B neutrinos are converted into another active flavor.
We have shown that the low-$T$ model 
is excluded by more than 5 $\sigma$ (3 $\sigma$) with data of 
the three (two out of the three) experiments. 
The best fitted value of $^{7}$Be neutrino flux is always negative 
even if we do not impose the luminosity constraint. 

On the other hand, if we assume that neutrino flavor conversion 
of $\nu_e \to \nu_\mu$ or $\nu_\tau$ is occurring, 
the best fitted flux of $^{7}$Be (or intermediate energy) neutrino 
is no longer negative. The current solar neutrino data suggest, 
as the best fit in our analysis, that 
($\langle {P_B} \rangle,  \langle {P_I} \rangle,   
\langle {P_{pp}} \rangle$) 
$\sim $(0.4,  0.2,  0.8).  
While it is still suppressed the value of the intermediate 
energy neutrinos makes most notable difference between 
cases with and without neutrino flavor conversion. We hope 
that this point is resolved by the future solar neutrino 
experiments.

More detailed discussions and relevant references are 
found in ref.\ [{21}]. 

\section*{ Acknowledgements}
This work was triggered by enlighting comments by Kenzo Nakamura 
at Hachimantai meeting in October 28-30, 1997, Iwate, Japan, 
to whom we are grateful. 
H. M. is partly supported by Grant-in-Aid for Scientific Research
Nos. 09640370 and 10140221 of the Japanese Ministry of Education, 
Science and Culture, and by Grant-in-Aid for Scientific Research 
No. 09045036 under the International Scientific Research Program, 
Inter-University Cooperative Research.
H. N. has been supported by a postdoctoral fellowship from Brazilian 
funding agency FAPESP. 
H. N. thanks A. Rossi for useful discussion. 
\vglue -2cm
\section*{References}
\re  
1.  
Arpesella, C. {\it et al.}, BOREXINO proposal, Vols. 1 and 2, 
ed. by Bellin, G. {\it et al.} (University of Milano, Milano, 1992);
Raghavan, S., 1995, Science {\bf 267}, 45.

\re 
2. 
Bahcall, J. N., 1989 {\it Neutrino Astrophysics}, 
Cambridge Univ. Press, Cambridge. 

\re  
3. 
Bahcall, J. N.  and Bethe, H. A., 1990, {\it Phys.\ Rev.\ Lett.}
{\bf 65}, 2233.

\re  
4. 
Bahcall, J. N., 1994, {\it Phys. Lett.} {\bf B338}, 276. 

\re  
5. 
Bahcall, J. N.  and  Krastev, P. I., 1996, {\it Phys. Rev.} 
{\bf D53}, 4211. 

\re  
6. 
Bahcall, J. N. and Ulmer, A., 1996, {\it Phys. Rev.} {\bf D53}, 4202. 

\re  
7.
Bahcall, J. N., Basu,  S.  and Pinsonneault,  M.H.,  
1998, {\it Phys. Lett.} B {\bf 433}, 1.

\re  
8.    
Bahcall, J. N., Krastev  P. I. and Smirnov, A. Yu.,  
1998, {\it Phys. Rev.} {\bf D58}, 096016.

\re  
9. 
Bandler, S.R. {\it et al.}, 1993, J. Low Temp. Phys. {\bf 93}, 785; 
Lanou, R.E., Maris, H.J. and Seidel, G.M.,  
1987, {\it Phys. Rev. Lett.} {\bf 58}, 2498. 

\re  
10. 
Bludman, S. , Kenneday, D.  Hata,  N.  and Langacker,  P.,  1993, 
{\it Phys. Rev.} {\bf D47}, 2220;
Castellani, V., Degl'Innocenti, S., Fiorentini,  G.,  
1993, {\it Astron. Astrophys.} {\bf 271}, 601;
Berezinsky, V., 1994 {\it Comm. Nucl. Part. Phys.} {\bf 21}, 249;
Degl'Innocenti, S., Fiorentini, G. and Lissia, M.,  
1995, {\it Nucl. Phys. Proc. Suppl.} {\bf 43}, 66;
Castellani, V. {\it et. al.}, 1997, {\it Phys. Rep.} {\bf 281}, 309;
Castellani, V., Degl'Innocenti  S. and Fiorentini, G., 1993,  
{\it Phys. Lett.} {\bf B303}, 68. 

\re  
11.  
Cleveland, B. T. {\it et al.}, 1998, {\it Astrophys.\ J. } 
{\bf 496}, 505;
Lande, K.\ (Homestake Collaboration), 
Talk given at {\it XVIII International Conference on Neutrino
Physics and Astrophysics (Neutrino '98)}, June, 1998, Takayama, Japan,  
to appear in the proceedings. 

\re  
12. 
Ewan, G. T.,  {\it et al.}, SNO collaboration, 
Sudbury Neutrino Observatory Proposal, SNO-87-12 (1987). 

\re 
13. 
M. Fukugita, Talk at Oji International Seminar on Elementary Processes 
in Dense Plasma, Tomakomai, Japan, June 27-July 1, 1994; preprint 
YITP-K-1086, August, 1994.

\re  
14. 
Gavrin, V.\ N.\  (SAGE Collaboration) in {\em Neutrino '98} [{11}].

\re  
15. 
Hata, N., Bludman,  S. and Langacker,  P., 1994 
{\it Phys. Rev.} {\bf D49}, 3622. 

\re  
16. 
Hata, N.  and Langacker, P., 1995, {\it Phys. Rev.} {\bf D52}, 420. 

\re 
17. 
Heeger, K. H.  and Robertson, R.\ G.\ H.\ , 
1996, {\it Phys. Rev. Lett.} {\bf 77}, 3270. 

\re  
18. 
Kristen, T.\  (GALLEX Collaboration), 
in {\em Neutrino '98} [{11}]. 

\re  
19. 
Kwong, W. and Rosen,  S. P., 1994, 
{\it Phys. Rev. Lett},  {\bf 73}, 369.

\re  
20. 
Laurenti, G. {\it et al.}, in {\it Proceedings of the Fith 
International Workshop on Neutrino Telescopes}, Venice, 
Italy, 1993, edited by M. Baldo Ceolin (Padua University, 
Padua, Italy, 1994), p. 161; 
Bonvicini, G., 1994, {\it Nucl. Phys.} {\bf B35}, 438. 

\re %
21. 
Minakata, H. and Nunokawa,  H., 1998, hep-ph/9810387. 

\re 
22. 
Spiro, M. and Vignaud,  D., 1990, 
{\it Phys. Lett.} {\bf B242}, 279.

\re  
23. 
Suzuki, Y.  (SuperKamiokande Collaboration), in {\em Neutrino '98} [{11}].

\re 
24. 
Parke, S., 1995,  {\it Phys. Rev. Lett},  {\bf 74}, 839. 

\re  
25. 
Smirnov, A. Yu. in {\it the proceedings of
XVII International conference on Neutrino Physics and Astrophysics, 
(Neutrino '96)} Helsinki, Finland, 1996, June, p38.

\end{document}